\documentclass[12pt,a4paper]{article}
\usepackage{amsmath}
\usepackage{amssymb}
\usepackage{hyperref}
\usepackage{graphicx}
\usepackage{multirow}
\usepackage{float}
\usepackage{epsfig}
\usepackage{array}
\usepackage[numbers,sort&compress]{natbib}
\title{This is your title}
\author{Nazanin Shiri}
\begin{document}
\date{\small\textsl{\today}}
	\title{\textsl Bound state solutions of the Schrödinger	equation for dibaryons via asymptotic iteration method}
 \author{\large
N.Shiri\footnote{Corresponding author.{\em E-mail addresses:}
nzn.shiry@grad.kashanu.ac.ir },\
\large N.Tazimi\footnote{{\em E-mail addresses:}
	tazimi@kashanu.ac.ir},
\large M.Monemzadeh\footnote{{\em E-mail addresses:}
	monem@kashanu.ac.ir}
	$\vspace{.5cm}$  \\\
\small{\em Department of Particle Physics, University of Kashan,
Kashan, Iran}\vspace{-1mm}}\maketitle
\vspace{.9cm} 
\begin{abstract}
Conventionally, hexaquarks are claimed to be exotic particles, most of which have not yet been experimentally detected. In this work, we study the mass spectra of exotic hadrons known as hexaquarks in the form of dibaryons. We investigate the hexaquark states with the two-body configuration in more detail.\\
 Starting from the analytical solution of the radial Schr\"odinger equation for the Hulth\'en potential in the framework of the asymptotic iteration method (AIM), we obtain the binding energy and mass spectrum of charm and bottom hexaquarks for different spin states. We strongly recommend searching experimentally for double charm and bottom dibaryons in the future.
\end{abstract}
\vspace{.5cm}
\section{Introduction} \label{introduction}
Since the inception of the quark-parton model and quantum chromodynamics (QCD), hadrons with unusual quantum numbers and multiquark fractions have attracted the interest of physicists. Conventional hadrons have a quark-antiquark (meson) or three-quark (baryon) composition. Unusual or exotic hadrons are expected to consist of four or more valence quarks or to contain valence gluons. The main reason for the intensive study of four-quark states was a mass hierarchy within the lowest scalar multiplet, which found its explanation in the framework of the four-quark model proposed by R. Jaffe \cite{1}. From 2003, i.e. from the first observation of the exotic meson X(3872), theoretical and experimental studies of tetra- and pentaquarks became one of the most interesting and fast-growing branches of high-energy physics. The valuable experimental information gathered in recent years, together with the theoretical progress achieved so far, now form two essential components of the physics of exotic hadrons \cite{2,3,4,5,6}.\\
Theoretically, the possibility of the existence of dibaryon states was first proposed by Dyson and Xuong in 1964 based on $SU(6)$ symmetry \cite{7}. Since then, extensive efforts have been made to explore the possible existence of a $\Delta\Delta$-dibaryon with hadronic degrees of freedom. However, no convincing results have been published so far. Since the birth of the quark model, dramatic progress has been made in this regard. In the following years, by including the interaction between the quark field and the chiral field in the constituent quark model, it was possible to reproduce the nucleon-nucleon (NN) interaction data and the binding energy of the deuteron \cite{8}, which would provide a much more reliable platform for predicting the structures of dibaryons in the quark degrees of freedom.\\
In recent decades, resonance structures with properties that cannot be described by normal mesons or baryons have been observed in several experiments at LHCb, WASA-at-COSY, BELLE, and elsewhere. In these experiments, states consisting of four quarks (tetraquarks), five quarks (pentaquarks), or six quarks (hexaquarks) have been observed \cite{9,10,11}. Jaffe first found the $H$ particle whose hyperfine interaction is much larger than that for two separate $\Lambda$-baryons in the chromomagnetic interaction model \cite{1}, and this dibaryon $uuddss$ has been studied in other frameworks as well \cite{12,13,14}. Another dibaryon candidate is the $d^*(2380)$, its quantum numbers $I(J^P) = 0(3^+)$, observed by the WASA-at-COSY Collaboration \cite{15}. In addition, the heavy dibaryons $qqqqqQ$ \cite{16}, the double heavy dibaryons $qqqqQQ$ \cite{17,18}, the triple heavy dibaryons $qqqQQQ$ \cite{19,20}, the other quite heavy dibaryons $QQQQQQ$ \cite{21}, and even the quite light dibaryons $qqqqqq$ \cite{22} have also been discussed. For a fully heavy system, since the constituent quarks are heavy, the relativistic effects are negligible and the kinetic energy is small. Lyu et al. in Ref. \cite{23} used lattice QCD to study the $\Omega_{ccc}\Omega_{ccc}$ in the $^1 S_0$ channel. They found this system is loosely bound by about 5.68 MeV. Within the same methodology, Mathur et al. replaced the charm quarks with the bottom quarks, and found a very deeply bound $\Omega_{bbb}\Omega_{bbb}$ dibaryon in the same channel, with a binding energy of about 89 MeV \cite{24}. The hadronic states, consisting of three quarks and three antiquarks, are another class of hexaquarks. The hidden-charm and hidden-bottom hexaquarks are particularly in focus because they have much larger masses and are therefore easier to distinguish from ordinary mesons. With the tetraquark and pentaquark states with hidden charms observed in the experiment, the discovery of hexaquarks with the hidden charm would also come true in the future. In Ref. \cite{25} we have calculated the mass spectrum of heavy tetraquarks, which is another verification of the existence of heavy exotic hadrons. \\ 
In recent decades BESIII collaboration measured the cross-section of action of the process $e^+e^-\rightarrow\pi^+\pi^-\psi$ (3686) and confirmed the existence of three charmonium-like states, with Y(4660) close to the threshold of $\Lambda_c\bar{\Lambda_c}$ systems \cite{26}. Previously, the Y(4660) structure was observed in the $e^+e^-\rightarrow\gamma_{ISR}\pi^+\pi^-\psi$ (3686) process in the Belle and BABAR experiments \cite{27,28}. Y(4630) was produced in the process $e^+e^-\rightarrow\Lambda_c\bar{\Lambda_c}$ in the Belle experiments \cite{29} and is considered to be a candidate for $\Lambda_c\bar{\Lambda_c}$ bound state \cite{30}. In particular, chiral perturbation theory for heavy baryons has been applied to systematically study the $\Lambda_c\bar{\Lambda_c}$, $\Sigma_c\bar{\Sigma_c}$, and $\Lambda_b\bar{\Lambda_b}$ systems \cite{31}, and the results suggest that Y(4260) and Y(4360) may be $\Lambda_c\bar{\Lambda_c}$ baryonia.\\
In this study, we have considered the hexaquark as a dibaryon because the study of the two-body systems is more convenient than a six-body system. For this purpose, we used the radial Schroedinger equation. The energy eigenvalues and the corresponding eigenfunctions between interaction systems have attracted much interest in both relativistic quantum mechanics and nonrelativistic quantum mechanics. The exact solution of the wave equations (relativistic or non-relativistic) is very important because the wave function contains all the necessary information about the quantum system under consideration. Analytical methods such as the supersymmetry method (SUSY) \cite{32} and the Nikiforov-Uvarov method (NU) \cite{33} have been used to solve the wave equations for quantum numbers with nonzero angular momentum $l=0$ using a given potential exactly or quasi-exactly. The radial Schr\"odinger equation for the Hulth\'en potential is solved exactly for $l\neq0$ using various methods \cite{34}. For the case that $l\neq0$, the effective Hulth\'en potential cannot be solved exactly, but there are various methods to find the eigenvalues of the energy in the bound state numerically \cite{35} and quasi-analytically. The Hulth\'en potential \cite{34} is one of the most important short-range potentials in physics and has been applied in various fields, including nuclear and particle physics, atomic physics, condensed matter, and chemical physics.\\
This paper is organized as follows: In the second section we thoroughly introduce the AIM method, then in section \ref{sec3} we use the Hulth\'en potential to obtain the energy eigenvalues and the corresponding eigenfunctions for arbitrary $l$-states. In section \ref{sec4} we solve the full spin effects terms for various $L$ and $S$ states. Then, in section \ref{sec5}, we  obtain the total mass spectrum of hexaquarks. In the last section we give a preliminary summary of our work.
\section{Asymptotic Iteration Method}\label{sec2}
\subsection{Energy Eigenvalues}
Homogeneous linear differential equations of second order occur naturally in many areas of mathematical physics. Many techniques can be found in the literature that can be used to solve this type of differential equation with boundary conditions. The main task of the present section is to introduce a new technique, which we call the asymptotic iteration method, for solving homogeneous second-order linear differential equations of the form
\begin{equation}\label{e1}
y^{\prime\prime}=\lambda_0(x)y^\prime+s_0(x)y
\end{equation}
where prime denotes the derivative concerning $x$. $\lambda_0(x)$ and $s_0(x)$ are defined in an interval that is not necessarily bounded, and $\lambda_0(x)$ and $s_0(x)$ have sufficiently many continuous derivatives and $\lambda_0(x)\neq0$. Let us consider the homogeneous second order linear differential Eq.\ref{e1} while $\lambda_0(x)$ and $s_0(x)$ are functions in $C_{\infty}(a,b)$. To find a general solution to this equation, we consider the symmetric structure of the right-hand side of Eq.\ref{e1}. The variables $s_0(x)$
and $\lambda_0(x)$ are sufficiently differentiable. Indeed, if we differentiate Eq.\ref{e1} with respect to $x$, we find that
\begin{equation}\label{e2}
y^{\prime\prime\prime}=\lambda_1(x)y^\prime+s_1(x)y
\end{equation}
where
\begin{equation}\label{e3}
\lambda_1=\lambda^\prime_0+s_0+\lambda_0^2 \qquad and \qquad s_1=s^\prime _0+s_0\lambda_0.
\end{equation}
AIM is briefly outlined here and the details can be found in \cite{36}. The differential Eq.\ref{e1} has a general solution \cite{36}
\begin{equation}\label{e4}
\begin{split}
y(x)&=\exp\left(-\int^{x}\alpha(x_1)\,\mathrm{d}x_1\right)\\
&\times\left[C_2+C_1\int^{x}exp\left(\int^{x_1}[\lambda_0(x_2)+2\alpha(x_2)]\,\mathrm{d}x_2\right)\,\mathrm{d}x_1\right]
\end{split}
\end{equation}
for $k>0$ and sufficiently large $k$, if
\begin{equation}\label{e5}
\frac{s_k(x)}{\lambda_k(x)}=\frac{s_{k-1}(x)}{\lambda_{k-1}(x)}=\alpha(x),  \qquad  k=1,2,3,...,
\end{equation} 
where
\begin{equation}\label{e6}
\begin{split}
\lambda_k& =\lambda^\prime_{k-1}(x)+s_{k-1}(x)+\lambda_0(x)\lambda_{k-1}(x), \\
s_{k}(x)&=s^\prime_{k-1}(x)+s_0(x)\lambda_{k-1}(x),    \qquad k=1,2,3,...
\end{split}
\end{equation}
Note that one can also start the recursion relations from $k=0$ with initial conditions $\lambda_1=1$ and $s_{-1}=0$ \cite{37}.
For a given potential, such as the Hulth\'en potential, the radial Schr\"odinger equation is written in the form of Eq.\ref{e1}. Then $s_0(x)$ and $\lambda_0(x)$ are determined and the parameters $s_k(x)$ and $\lambda_k(x)$ are obtained by the recurrence relations given in Eq.\ref{e6}. The termination condition of the method in Eq.\ref{e5} can be designed as follows
\begin{equation}\label{e7}
\Delta_k(x)=\lambda_k(x)s_{k-1}(x)-\lambda_{k-1}(x)s_k(x)=0,  \qquad  k=1,2,3,...,
\end{equation}
The energy eigenvalues are obtained from the roots of Eq.\ref{e7} if the problem is exactly solvable. If not, a suitable $n$ point is chosen for a given $x_0$ principal quantum number, generally determined as the maximum value of the asymptotic wavefunction or the minimum value of the potential \cite{36,37}, and the approximate energy eigenvalues are obtained from the roots of this equation for sufficiently large values of $k$ with iteration.
\subsection{Energy Eigenfunctions}
In this study, we seek the exact solution of the radial Schr\"odinger equation, for which the relevant second-order homogeneous linear differential equation has the following general form \cite{36}
\begin{equation}\label{e8}
y^{\prime\prime}=2\left(\frac{ax^{N+1}}{1-bx^{N+2}}-\frac{(t+1)}{x}\right)y^\prime-\frac{\omega_k^t(N)x^N}{1-bx^{N+2}}y,     \qquad 0<x<\infty.
\end{equation}
If this equation is compared to Eq.\eqref{e1}, it entails the following expressions
\begin{equation}\label{e9}
\begin{split}
\lambda_0(x)&=2\left(\frac{ax^{N+1}}{1-bx^{N+2}}-\frac{(t+1)}{x}\right),\\
s_0(x)&=-\frac{\omega_k^t(N)x^N}{1-bx^{N+2}}.
\end{split}
\end{equation}
$a$ and $b$ are constants and $\omega_k^t(N)$ can be determined from condition Eq.\ref{e5} for $k = 0, 1, 2, 3, \dots$ and $N = -1, 0, 1, 2, 3, \dots$ as follows
\begin{align*}
\omega_k^t(-1)&=k(2a+2bt+(k+1)b),\\
\omega_k^t(0)&=2k(2a+2bt+(2k+1)b),\\
\omega_k^t(1)&=3k(2a+2bt+(3k+1)b),\\
\omega_k^t(2)&=4k(2a+2bt+(4k+1)b),\\
\omega_k^t(3)&=5k(2a+2bt+(5k+1)b),\\
\vdots \notag
\end{align*}
Hence, these formulae are easily generalized as
\begin{equation}\label{e10}
\omega_k^t(N)=b(N+2)^2k\left(k+\frac{(2t+1)b+2a}{(N+2)b}\right).
\end{equation}
The exact eigenfunctions can be derived from the following generator
\begin{equation}\label{e11}
y_n(x)=C_2exp\left(-\int^x\frac{s_k(x^\prime)}{\lambda_k(x^\prime)}\,\mathrm{d}x^\prime\right),
\end{equation}
where $k\geq n$, $k$ is the iteration number and $n$ is the radial quantum number. For exactly solvable potentials, the iteration number is equal to the radial quantum number ($n=k$) and the eigenfunctions follow directly from Eq.\ref{e11}. For nontrivial potentials that do not have exact solutions, in this numerical solution, $k$ is always larger than $n$ and the approximated energy eigenvalues are obtained from the root of Eq.\ref{e7} for sufficiently large values of $k$ with iteration. It should be noted that $\alpha_x$  calculated from Eq.\ref{e5} is zero for the ground state. Thus, if you replace Eq.\ref{e5} with Eq.\ref{e9} in Eq.\ref{e11}, the eigenfunction is obtained. Finally, the following general formula for the exact solutions $y_n(x)$ is obtained as
\begin{equation}\label{e12}
y_n(x)=(-1)^nC_2(N+2)^n(\sigma)_n \, _2F_1(-n,\rho+n;\sigma;bx^{(N+2)}).
\end{equation}
It is important to note that the square integrable in $L^2$ is this total wavefunction, which is the asymptotic form of the wavefunction times $y_n(x)$ given by Eq.\ref{e11}. Here
\begin{align*}
(\sigma)_n&=\frac{\Gamma(\sigma+n)}{\Gamma(\sigma)},\\
\sigma&=\frac{2t+N+3}{N+2},\\
\rho&=\frac{(2t+1)b+2a}{(N+2)b}.
\end{align*}
$\sigma_n$ and $_2F_1$ are known as the Pochhammer symbol and the Gauss hypergeometric function, respectively.
\section{Eigenvalues and Eigenfunctions}\label{sec3}
The motion of a particle with the mass $M$ in the spherically symmetric potential is described in the spherical coordinates by the following Schr\"odinger equation
\begin{equation}\label{e13}
\begin{split}
\frac{-\hbar^2}{2M}&\left(\frac{\partial^2}{\partial r^2}+\frac 2r\frac{\partial}{\partial r}+\frac{1}{r^2}\left[\frac{1}{\sin\theta}\frac{\partial}{\partial\theta}\left(\sin\theta\frac{\partial}{\partial\theta}\right)+\frac{1}{\sin^2\theta}\frac{\partial^2}{\partial\phi^2}\right]+V(r)\right)\\
&\times\Psi_{nlm}(r,\theta,\phi)=E\Psi_{nlm}(r,\theta,\phi).
\end{split}
\end{equation}
By substituting $\Psi_{nlm}(r,\theta,\phi)=R_{nl}(r)Y_{lm}(\theta,\phi)$ , the radial part of \\ the schr\"odinger equation becomes
\begin{equation}\label{e14}
\left(\frac{\mathrm{d}^2}{\mathrm{d}r^2}+\frac{2}{r}\frac{\mathrm{d}}{\mathrm{d}r}\right)R_{nl}(r)+\frac{2M}{\hbar^2}\left[E-V(r)-\frac{l(l+1)\hbar^2}{2Mr^2}\right]R_{nl}(r)=0.
\end{equation}
It is sometimes convenient to define $R_{nl}(r)$ and the effective potential as follows
\begin{equation}\label{e15}
R_{nl}(r)=\frac{u_{nl}(r)}{r},  \qquad V_{eff}=V(r)+\frac{l(l+1)\hbar^2}{2Mr^2}.
\end{equation}
Since
\begin{equation}\label{e16}
\left(\frac{\mathrm{d}^2}{\mathrm{d}r^2}+\frac{2}{r}\frac{\mathrm{d}}{\mathrm{d}r}\right)\frac{u_{nl}(r)}{r}=\frac{1}{r}\frac{\mathrm{d}^2}{\mathrm{d}r^2}u_{nl}(r),
\end{equation}
the radial Schr\"odinger equation \cite{35} given by Eq.\ref{e14} follows that
\begin{equation}\label{e17}
\frac{\mathrm{d}^2u_{nl(r)}}{\mathrm{d}r^2}+\frac{2M}{\hbar^2}[E-V_{eff}]u_{nl}(r)=0.
\end{equation}
In this section, we shall introduce the Hulth\'en potential to solve the radial Schr\"odinger equation. The Hulth\'en potential \cite{34} is given by
\begin{equation}\label{e18}
V_H(r)=-Ze^2\delta\frac{e^{-\delta r}}{1-e^{-\delta r}},
\end{equation}
where $Z$ and $\delta$ are the atomic numbers and the screening parameter, respectively, which determine the range for the Hulth\'en potential. The Hulth\'en potential behaves like the Coulomb potential near the origin $(r\rightarrow 0)$, but in the asymptotic region $(r\gg1)$ the Hulth\'en potential decreases exponentially so that its capacity for bound states is smaller than that of the Coulomb potential. However, for small values of the screening parameter $\delta$, the Hulth\'en potential becomes the Coulomb potential given by $V_C=\frac{-Ze^2}{r}$. The effective Hulth\'en potential is
\begin{equation}\label{e19}
V_{eff}(r)=V_H(r)+V_l=-Ze^2\delta\frac{e^{-\delta r}}{1-e^{-\delta r}}+\frac{l(l+1)\hbar^2}{2Mr^2},
\end{equation}
where $V_l=\frac{l(l+1)\hbar^2}{2Mr^2}$ is known as the centrifugal term. This effective potential cannot be solved analytically for $l\neq0$ because of the centrifugal term. Therefore, we must use an approximation for the centrifugal term, similar to other authors \cite{38,39}. In this approximation, $\frac{1}{r^2}=\delta^2\frac{e^{-\delta r}}{(1-e^{-\delta r})^2}$ is used for the centrifugal term. This is valid only for small $\delta r$ and breaks down in the high screening region. For small $\delta r$, $\tilde{V}_{eff}(r)$ is very well approximated to ${V}_{eff}(r)$ and the schr\"odinger equation for this approximated potential is analytically solvable. The effective potential is thus
\begin{equation}\label{e20}
\tilde{V}_{eff}(r)=-Ze^2\delta\frac{e^{-\delta r}}{1-e^{-\delta r}}+\frac{l(l+1)\hbar^2\delta^2}{2M}\frac{e^{-\delta r}}{(1-e^{-\delta r})^2},
\end{equation}
Instead of using the radial Schr\"odinger equation for the effective Hulth\'en potential ${V}_{eff}(r)$ according to Eq.\ref{e19}, we now solve the radial Schr\"odinger equation for the new effective potential $\tilde{V}_{eff}(r)$ according to Eq.\ref{e20}. Put this new effective potential into
Eq.\eqref{e17} and use the following ans\"atze equation to make the differential equation more compact
\begin{equation}\label{e21}
-\varepsilon^2=\frac{2ME}{\hbar^2\delta^2}, \qquad  \beta^2=\frac{2MZe^2}{\hbar^2\delta}, \qquad \delta r=x,
\end{equation}
and if we rewrite the radial Schr\"odinger equation by using a new variable of the form $z=e^{-x}$, we obtain
\begin{equation}\label{e22}
\frac{\mathrm{d}^2u_{nl}(z)}{\mathrm{d}z^2}+\frac{1}{z}\frac{\mathrm{d}u_{nl}(z)}{\mathrm{d}z}+\left[-\frac{\varepsilon^2}{z^2}+\frac{\beta^2}{z(1-z)}-\frac{l(l+1)}{z(1-z)^2}\right]u_{nl}(z)=0.
\end{equation}
In order to solve this equation with AIM, we should transform this equation to the form of Eq.\ref{e1}. Therefore, the reasonable physical wavefunction we propose is as follows
\begin{equation}\label{e23}
u_{nl}(z)=z^{\varepsilon}(1-z)^{l+1}f_{nl}(z).
\end{equation}
If we insert this wavefunction into Eq.\ref{e22}, we have the second-order homogeneous linear differential equations in the following form 
\begin{equation}\label{e24}
\begin{split}
\frac{\mathrm{d}^2f_{nl}(z)}{\mathrm{d}z^2}&=\left[\frac{(2\varepsilon+2l+3)z-(2\varepsilon+1)}{z(1-z)}\right]\frac{\mathrm{d}f_{nl}(z)}{\mathrm{d}z}\\
&+\left[\frac{(2\varepsilon+l+2)l+2\varepsilon-\beta^2+1}{z(1-z)}\right]f_{nl}(z),
\end{split}
\end{equation}
which is now amenable to an AIM solution. By comparing this equation with Eq.\ref{e1}, we can write the $\lambda_0(z)$ and $s_0(z)$ values and by means of Eq.\ref{e6} we may calculate $\lambda_k(z)$ and $s_k(z)$. By combining the results with the quantization given by Eq.\ref{e7} for different $k=0,1,2,\dots$ and using Eq.\ref{e21} we will obtain the energy eigenvalues $E_{nl}$,
\begin{equation}\label{e25}
E_{nl}=-\frac{\hbar^2}{2M}\left[\frac{MZe^2}{\hbar^2(n+l+1)}-\frac{(n+l+1)\delta}{2}\right]^2.
\end{equation}
In the atomic units $(\hbar=e=1)$ and for $Z=1$, Eq.\ref{e25} turns out to be
\begin{equation}\label{e26}
E_{nl}=-\frac{1}{2M}\left[\frac{M}{(n+l+1)}-\frac{(n+l+1)\delta}{2}\right]^2.
\end{equation}
Here $M$ is the reduced mass of baryons.\\
Now, as indicated in Sec.\ref{sec2}, we can determine the corresponding wavefunctions by using Eq.\ref{e12}. When we compare Eq.\ref{e8} and Eq.\ref{e24}, we find $N=-1$, $b=1$, $a=l+1$, and $t=\frac{2\varepsilon-1}{2}$. Therefore, we find $\rho=2(\varepsilon+l+1)$ and $\sigma=2\varepsilon+1$. So we can easily find the solution for $f_{nl}(z)$ for the energy eigenvalue Eq.\ref{e25} by using Eq.\ref{e12} 
\begin{equation}\label{e27}
f_{nl}(z)=(-1)^n\frac{\Gamma(2\varepsilon_n+n+1)}{\Gamma(2\varepsilon_n+1)} \, _2F_1(-n,2\varepsilon_n+2l+2+n;2\varepsilon_n+1;z).
\end{equation}
Thus, we can write the total radial wavefunction as follows:
\begin{equation}\label{e28}
u_{nl}(z)=Nz^{\varepsilon_n}(1-z)^{l+1} \, _2F_1(-n,2(\varepsilon_n+l+1)+n;2\varepsilon_n+1;z),
\end{equation}
where N is the normalization constant.
\section{Spin Effects on Potential} \label{sec4}
\subsection{Whole Terms of Spin Effects}
Thus, the present section aims at re-examining the above configuration in a full-fledged calculation taking into account the spin effect on the mass of the dibaryonic particles.
Just as we calculated the mass spectrum of heavy tetraquarks including the spin effect in Ref.\cite{25}, we use the same method here to include the spin effect in the mass calculations as well.\\
Assuming that our two-body system resembles a tetraquark system, the contribution of spin-dependent potentials, a spin-spin $V_{ SS }(r)$, spin-orbit $V_{ LS }(r)$, and tensor $V_{T}(r)$, which gives significant contributions, especially for excited states, is necessary to better understand the partitioning between orbital and radial excitations of different combinations of quantum numbers of dibaryons. All three spin-dependent terms are driven by the Breit-Fermi Hamiltonian for the one-gluon exchange \cite{40,41}, yielding
\begin{equation}\label{e29}
V_{SS}(r)=C_{SS}(r)S_1.S_2,
\end{equation}
\begin{equation}\label{e30}
V_{LS}(r)=C_{LS}(r)L.S,
\end{equation}
\begin{equation}\label{e31}
V_{T}(r)=C_{T}(r)S_{12},
\end{equation}
The matrix element $S_1.S_2$ acts on the wave function and produces a constant factor, but $V_{ SS }(r)$ remains a function only of $r$, and the expectation values of $\left<S_1.S_2\right>$ are available by a quantum mechanical formula \cite{42}.
\begin{equation}\label{e32}
\left<S_1.S_2\right>=\left<\frac 12(S^2-S_1^2-S_2^2)\right>, \qquad S=S_1+S_2
\end{equation}
where $S$, $S_1$ and $S_2$ denote the total spin and the spins of constituent baryons in dibaryons component, respectively. $C_{ SS }(r)$ may be defined as follows
\begin{equation}\label{e33}
C_{SS}(r)=\frac{2}{3m^2}\nabla^2V_V(r)=-\frac{8\kappa_s\alpha_s\pi}{3m^2}\delta^3(r),
\end{equation}
A fair match can be obtained by adding the spin-spin interaction in a zero-order potential using the Schr\"odinger equation in dibaryon spectroscopy by including the spin-spin interaction using the artifact that gives a new parameter $\sigma$ instead of the Dirac delta. Thus, $V_{ SS }(r)$ can now be redefined as
\begin{equation}\label{e34}
V_{SS}(r)=-\frac{8\kappa_s\alpha_s\pi}{3m^2}(\frac{\sigma}{\sqrt{\pi}})^3\exp{(-\sigma^2(r)^2)}S_1.S_2,
\end{equation}
The expected value of the operator $\left<L.S\right>$ depends mainly on the total angular momentum $J$, which is calculated according to the formula $J=L+S$,
\begin{equation}\label{e35}
\left<L.S\right>=\left<\frac12(J^2-L^2-S^2)\right>\equiv\frac12\left[J(J+1)-S(S+1)-l(l+1)\right]
\end{equation}
where $L$ denotes the total orbital angular momentum of the quarks in the case of the dibaryon. The following equation can be used to calculate $C_{ LS }(r)$ :
\begin{equation}\label{e36}
C_{LS}(r)=-\frac{3\kappa_s\alpha_s\pi}{2m^2}\frac{1}{(r)^2}-\frac{c}{2m^2}\frac{1}{(r)}
\end{equation}
The second component of the spin-orbit interaction is called the Thomas precession and is proportional to the scalar term. $c=0.2 \ GeV^2$ is the fitting parameter. The confining interaction is thought to be due to the Lorentz scalar structure. Thus, $V_{ LS }(r)$ can now be redefined as
\begin{equation}\label{e37}
V_{LS}(r)=\left[-\frac{3\kappa_s\alpha_s\pi}{2m^2}\frac{1}{(r)^2}-\frac{c}{2m^2}\frac{1}{(r)}\right]\left< L.S \right>.
\end{equation}
 In higher excited states, the contribution of the spin-tensor becomes quite important, which requires a little algebra and can be calculated as follows
\begin{equation}\label{e38}
C_{T}(r)=-\frac{12\kappa_s\alpha_s\pi}{4m^2}\frac{1}{(r)^3}
\end{equation}
The results of $(S_1.S_2)$ are obtained by solving the diagonal matrix elements for the particles with spin $\frac12$ and spin $1$, as described in the Ref.\cite{43}. To solve the tensor interaction, the simpler formulation can be used:
\begin{equation}\label{e39}
S_{12}=12(\frac{(S_1.(r))(S_2.(r))}{(r)^2}-\frac13(S_1.S_2))
\end{equation}
which can be redefined as
\begin{equation}\label{e40}
S_{12}=4[3(S_1.\hat{(r)})(S_2.\hat{(r)})-(S_1.S_2)]
\end{equation}
Pauli matrices and spherical harmonics with their corresponding eigenvalues can be used to obtain the results of the $S_{12}$ term. The following conclusions are valid for two-body systems:
\begin{equation}\label{e41}
\begin{split}
\left<S_{12}\right>_{\frac12\otimes\frac12\rightarrow S=1,l\neq0}&=\left\{ 
\begin{array}{lr}
-\frac{2l}{2l+3} \ for \ &J=l+1,\\ \\
-\frac{2(l+1)}{(2l-1)} \ for\ &J=l-1,\\ \\
2 \ for\  &J=l,
\end{array} \right.
\end{split}
\end{equation}
when $l=0$ and $S=0$, the $\left<S_{12}\right>$ always vanishes, but it yields a non-zero value for excited states
\begin{equation}\label{e42}
\begin{split}
\left<S_{12}\right>&=\left\{ 
\begin{array}{lr}
-\frac 25 \ for \ &J=2,\\ \\
+2 \ for\ &J=1,\\ \\
-4 \ for\  &J=0,
\end{array} \right.
\end{split}
\end{equation}
these values are valid only for particles with specific spin-half. All data were collected in Table\ref{t1},
where
\begin{equation}\label{e43}
V_{T}(r)=C_{T}(r)(\frac{(S_1.(r))(S_2.(r))}{(r)^2}-\frac13(S_1.S_2))
\end{equation}
The final and simple form of Eq.\ref{e43} is
\begin{equation}\label{e44}
V_{T}(r)=-\frac{12\kappa_s\alpha_s\pi}{4m^2}\frac{1}{(r)^3}\left<S_{12}\right>
\end{equation}
After a lengthy calculation, the following expression for the diagonal matrix elements of $\left<S_{12}\right>$ may be found \cite{44}:
\begin{equation}\label{e45}
\left<S_{12}\right>=\frac{4}{(2l+3)(2l-1)}\left[\left<S^2\right>\left<L^2\right>-\frac32\left<L.S\right>-3(\left<L.S\right>)^2\right]
\end{equation}
When the two-body problem is solved to obtain the masses of the dibaryon because the interaction between the three quarks inside the baryon is identical; when the S-wave state is considered, only the spin-spin interaction is relevant; the spin-orbit and the tensor are both identically zero for ground states. 
\subsection{Spherical Effective Potential}
According to Eq.\ref{e20} we have the effective potential. As asserted before in the atomic units $(\hbar=e=1)$ and for $Z=1$, we can rewrite the effective potential as follows:
\begin{equation}\label{e46}
\tilde{V}_{eff}(r)=-\delta\frac{e^{-\delta r}}{1-e^{-\delta r}}+\frac{l(l+1)\delta^2}{2\mu}\frac{e^{-\delta r}}{(1-e^{-\delta r})^2},
\end{equation}
where $\mu\equiv M\equiv m$ is the reduced mass of related dibaryon. In terms of the reduced mass,
let $\mu=\frac{m_1m_2}{m_1+m_2}$ where $m_1$ and $m_2$ are the constituent masses of baryon 1 and baryon 2, respectively. The contribution of spin-dependent terms can be calculated by writing total potential. The spin-spin correction to the non-relativistic potential can be obtained as \cite{45} 
\begin{equation}\label{e47}
V_{SS}(r)=\frac{8\pi\kappa_s\alpha_s}{3m^2}\int\psi^*(r)\psi(r)\delta(r)\left<S_1.S_2\right>\mathrm{d}^3r=\frac{8\pi\kappa_s\alpha_s}{3m^2}|\psi(0)|^2\left<S_1.S_2\right>.
\end{equation}
If the spin-spin interaction was treated as a first-order perturbation without the Gaussian smearing, as  is the case in this work, it would be proportional to modulus of the wavefunction at the origin, $|\psi(0)|^2$. Since spin-spin interaction only occurs in S-wave, only S-wave states (i.e., orbital angular momentum $l=0$) have non-zero value of the wavefunction at the origin. Therefore for S-wave state we have \cite{46}
\begin{equation}\label{e48}
|\psi(0)|^2=|Y_0^0(\theta,\phi)R_{n,l}(0)|^2=\frac{| R_{n,l}(0)|^2}{4\pi}.
\end{equation}
$|R_{n,l}(0)|^2$ can be obtained directly from the numerical calculations and is related to the radial potential as
\begin{equation}\label{e49}
|\psi(0)|^2=\frac{\mu}{2\pi}\left<\frac{d}{dr}V(r)\right>\Rightarrow|R_{n,l}(0)|^2=2\mu\left<\frac{d}{dr}V(r)\right>.
\end{equation}
So it is possible to replace $\lvert\psi(0)\rvert^2$ in Eq.\ref{e47} with $\lvert R_{n,l}(0)\rvert^2$ and by deriving Eq.\ref{e46} respect to $r$ we have
\begin{equation}\label{e50}
\left<\frac{d}{dr}\tilde{V}_{eff}(r)\right>=\delta^2e^{-\delta r}\left[\frac{l(l+1)}{2\mu}\delta(e^{-\delta r}-1)+1\right].
\end{equation}
The final form of spin-spin term for the potential is
\begin{equation}\label{e51}
V_{SS}(r)=\frac{8\pi\kappa_s\alpha_s}{3m^2}\frac{\mu}{2\pi}\delta^2e^{-\delta r}\left[\frac{l(l+1)}{2\mu}\delta(e^{-\delta r}-1)+1\right]\left<S_1.S_2\right>,
\end{equation}
where $\alpha_s$ is the QCD coupling constant. Choosing $\alpha_s=0.5$ constant is a common approach in many of the non-relativistic quark potential models. To calculate the dibaryon mass, we took the color factor as $\kappa_s=-\frac23$ (for baryon-baryon system). 
\begin{table}[H]
\caption{Expectation values for $\left<S_1.S_2\right>$, $\left<L.S\right>$ and $\left<S_{12}\right>$ in ground and exited states.}
\begin{center}
\begin{tabular}{c c c c c c }
\hline
S&L&J&$\left<S_1.S_2\right>$&$\left<L.S\right>$&$\left<S_{12}\right>$ \\ \hline \hline
\multicolumn{1}{c}{\multirow{3}{*}{1}}&0&1&\multicolumn{1}{c}{\multirow{3}{*}{$\frac14$}}&0&0 \\
\multicolumn{1}{c}{}&1&2&\multicolumn{1}{c}{}&1&$-\frac25$\\
\multicolumn{1}{c}{}&2&3&\multicolumn{1}{c}{}&2&$-\frac47$\\ \hline
\multicolumn{1}{c}{\multirow{3}{*}{2}}&0&2&\multicolumn{1}{c}{\multirow{3}{*}{$\frac34$}}&0&0\\
\multicolumn{1}{c}{}&1&3&\multicolumn{1}{c}{}&2&$-\frac{12}{5}$\\
\multicolumn{1}{c}{}&2&4&\multicolumn{1}{c}{}&4&$-\frac{24}{7}$\\ \hline
\multicolumn{1}{c}{\multirow{3}{*}{3}}&0&3&\multicolumn{1}{c}{\multirow{3}{*}{$\frac94$}}&0&0\\
\multicolumn{1}{c}{}&1&4&\multicolumn{1}{c}{}&3&-6\\
\multicolumn{1}{c}{}&2&5&\multicolumn{1}{c}{}&6&$-\frac{60}{7}$\\ 
\hline \hline
\end{tabular}
\end{center}
\label{t1}
\end{table}
To show the total spin effect, we have determined the expectation values of $\left<S_1.S_2\right>$, $\left<L.S\right>$ and $\left<S_{12}\right>$ for three spin terms $S=1,2,3$, into which we have inserted the orbital angular momentum $L=0,1,3$. As we see in Table\ref{t1}, for $L=0$ we have a value of zero for $\left<L.S\right>$ and $\left<S_{12}\right>$ since the orbital angular momentum effect acts directly on these terms, while there is a specific non-zero value for $\left<S_1.S_2\right>$ in each state of spin. As the spin and orbital angular momentum values increase, the value for $\left<S_{12}\right>$ decreases, perhaps because in the upper layer, the dependence on the tensor term is reduced.
\section{Total Mass of Hexaquarks}\label{sec5}
Exact or approximate solutions of the nonrelativistic radial schr\"odinger equation for two-body systems have attracted considerable attention. On the eve of the commissioning of the Wasa-at-Cosy and the search for new physics beyond the Standard Model, they were able to find a particle called $d^*(2380)$, an exotic particle in the form of a light hexaquark. Although there is still no complete laboratory data for these exotic six-quark particles, these studies have become a hot topic in particle physics. To study the general structure of the spectrum of exotic hexaquarks, we consider these structures as two-body systems called dibaryons. As described in Sec.\ref{sec3}, by solving the radial schr\"odinger equation with the iteration method, we have obtained the energy eigenvalue and the eigenfunction.\\
Now we apply Eq. \eqref{e52} to calculate the mass spectrum of hexaquark states. 
\begin{equation}\label{e52}
M_{Dibaryon}= M_{Baryon_1} + M_{Baryon_2} + E_{n,l}+\left< V_{SS}(r)\right>+\left< V_{LS}(r)\right>+\left< V_{T}(r)\right>
\end{equation} 
The terms we have calculated are shown in Table \ref{t2}; the first column is the quark content of the hexaquarks; the second column denotes the quantum number of the system; the third column is the corresponding binding energy state; the fourth, fifth, and sixth columns represent the spin-spin, orbit-spin, and tensor terms, respectively; and the last column represents the mass spectrum of the hexaquark.\\
\begin{table}
\caption{The values of hexaquarks subsystems for $S=1$, $n=1$, $L=0,1,2$, $\delta=0.25$ and $r=0.25$ are the total spin, the principal quantum number, screening and radial parameter, respectively. Quark content is $(q=u,d)$. The masses are all in units of MeV. All data come from PDG\cite{47}.}
\begin{center}
\begin{footnotesize}
\begin{tabular}{c c c c c c c}
\hline
Hexaquark&$J^{PC}$&$E_{nl}$&$V_{SS}$&$V_{LS}$&$V_{T}$&M\\ \hline \hline
\multicolumn{1}{c}{\multirow{3}{*}{$\Lambda_c\Lambda_c (qqqqcc)$}}&$1^{--}$&-143&$0.57\times10^{-5}$&0&0&4430\\
\multicolumn{1}{c}{}&$2^{++}$&-64&0.013&$0.19\times10^{-4}$&$-0.1\times10^{-4}$&4510\\
\multicolumn{1}{c}{}&$3^{--}$&-36&0.013&$0.38\times10^{-4}$&$-0.15\times10^{-4}$&4537\\ 

\multicolumn{1}{c}{\multirow{3}{*}{$\Sigma_c\Sigma_c (qqqqcc)$}}&$1^{--}$&-153&$0.53\times10^{-5}$&0&0&4754\\
\multicolumn{1}{c}{}&$2^{++}$&-68&0.013&$0.16\times10^{-4}$&$-0.9\times10^{-5}$&4840\\ 
\multicolumn{1}{c}{}&$3^{--}$&-38&0.013&$0.33\times10^{-4}$&$-0.13\times10^{-4}$&4869\\

\multicolumn{1}{c}{\multirow{3}{*}{$\Xi_c\Xi_c(qqsscc)$}}&$1^{--}$&-154&$0.52\times10^{-5}$&0&0&4788\\
\multicolumn{1}{c}{}&$2^{++}$&-69&0.013&$0.16\times10^{-4}$&$-0.87\times10^{-5}$&4873\\ 
\multicolumn{1}{c}{}&$3^{--}$&-39&0.013&$0.32\times10^{-4}$&$-0.1\times10^{-4}$&4903\\ 

\multicolumn{1}{c}{\multirow{3}{*}{$\Omega_c\Omega_c(sssscc)$}}&$1^{--}$&-168&$0.48\times10^{-5}$&0&0&5222\\
\multicolumn{1}{c}{}&$2^{++}$&-77&0.013&$0.136\times10^{-4}$&$-0.74\times10^{-5}$&5316\\ 
\multicolumn{1}{c}{}&$3^{--}$&-42&0.013&$0.27\times10^{-4}$&$-0.1\times10^{-4}$&5349\\ \\

\multicolumn{1}{c}{\multirow{3}{*}{$\Lambda_b\Lambda_b(qqqqbb)$}}&$1^{--}$&-351&$0.23\times10^{-5}$&0&0&10888\\
\multicolumn{1}{c}{}&$2^{++}$&-156&0.013&$0.3\times10^{-5}$&$-0.17\times10^{-5}$&11083\\
\multicolumn{1}{c}{}&$3^{--}$&-88&0.013&$0.6\times10^{-5}$&$-0.24\times10^{-5}$&11152\\ 

\multicolumn{1}{c}{\multirow{3}{*}{$\Sigma_b\Sigma_b(qqqqbb)$}}&$1^{--}$&-363&$0.22\times10^{-5}$&0&0&11258\\
\multicolumn{1}{c}{}&$2^{++}$&-161&0.013&$0.29\times10^{-5}$&$-0.16\times10^{-5}$&11460\\ 
\multicolumn{1}{c}{}&$3^{--}$&-91&0.013&$0.58\times10^{-5}$&$-0.23\times10^{-5}$&11531\\

\multicolumn{1}{c}{\multirow{3}{*}{$\Xi_b\Xi_b(qqssbb)$}}&$1^{--}$&-362&$0.22\times10^{-5}$&0&0&11222\\
\multicolumn{1}{c}{}&$2^{++}$&-161&0.013&$0.29\times10^{-5}$&$-0.16\times10^{-5}$&11423\\ 
\multicolumn{1}{c}{}&$3^{--}$&-90&0.013&$0.59\times10^{-5}$&$-0.23\times10^{-5}$&11494\\ 

\multicolumn{1}{c}{\multirow{3}{*}{$\Omega_b\Omega_b(ssssbb)$}}&$1^{--}$&-378&$0.21\times10^{-5}$&0&0&11713\\
\multicolumn{1}{c}{}&$2^{++}$&-168&0.013&$0.27\times10^{-5}$&$-0.14\times10^{-5}$&11923\\ 
\multicolumn{1}{c}{}&$3^{--}$&-94&0.013&$0.54\times10^{-5}$&$-0.21\times10^{-5}$&11996\\ \hline\hline
\end{tabular}
\end{footnotesize}
\end{center}
\label{t2}
\end{table}

\begin{table}[H]
\caption{The values of hexaquarks subsystems for $S=1$, $n=1$, $L=0,1,2$, $\delta=0.25$ and $r=0.25$ are the total spin, the principal quantum number, screening and radial parameter, respectively. The masses are all in units of MeV. All data come from PDG\cite{47}.  }
\begin{center}
\begin{small}
\begin{tabular}{c c c c c c c}
\hline
Hexaquark&$J^{PC}$&$E_{nl}$&$V_{SS}$&$V_{LS}$&$V_{T}$&M\\ \hline \hline
\multicolumn{1}{c}{\multirow{3}{*}{$\Sigma_c^{*}\Sigma_c$}}&$1^{--}$&-155&$0.5\times10^{-5}$&0&0&4817 \\
\multicolumn{1}{c}{}&$2^{++}$&-69&0.013&$0.16\times10^{-4}$&$-0.86\times10^{-5}$&4903\\
\multicolumn{1}{c}{}&$3^{--}$&-39&0.013&$0.3\times10^{-4}$&$-0.1\times10^{-4}$&4934\\

\multicolumn{1}{c}{\multirow{3}{*}{$\Xi_c^{*}\Xi_c$}}&$1^{--}$&-160&$0.5\times10^{-5}$&0&0&4958 \\
\multicolumn{1}{c}{}&$2^{++}$&-71&0.013&$0.15\times10^{-4}$&$-0.82\times10^{-5}$&5046\\
\multicolumn{1}{c}{}&$3^{--}$&-40&0.013&$0.3\times10^{-4}$&$-0.1\times10^{-4}$&5078\\

\multicolumn{1}{c}{\multirow{3}{*}{$\Omega_c^{*}\Omega_c$}}&$1^{--}$&-171&$0.47\times10^{-5}$&0&0&5291 \\
\multicolumn{1}{c}{}&$2^{++}$&-76&0.013&$0.13\times10^{-4}$&$-0.7\times10^{-5}$&5386\\
\multicolumn{1}{c}{}&$3^{--}$&-43&0.013&$0.26\times10^{-4}$&$-0.1\times10^{-4}$&5419\\

\multicolumn{1}{c}{\multirow{3}{*}{$\Sigma_b^{*}\Sigma_b$}}&$1^{--}$&-364&$0.23\times10^{-5}$&0&0&11277 \\
\multicolumn{1}{c}{}&$2^{++}$&-162&0.013&$0.29\times10^{-5}$&$-0.2\times10^{-5}$&11479\\
\multicolumn{1}{c}{}&$3^{--}$&-91&0.013&$0.6\times10^{-5}$&$-0.23\times10^{-5}$&11550\\ 

\multicolumn{1}{c}{\multirow{3}{*}{$\Xi_b^{*}\Xi_b$}}&$1^{--}$&-367&$0.22\times10^{-5}$&0&0&11380 \\
\multicolumn{1}{c}{}&$2^{++}$&-163&0.013&$0.29\times10^{-5}$&$-0.15\times10^{-5}$&11584\\
\multicolumn{1}{c}{}&$3^{--}$&-92&0.013&$0.6\times10^{-5}$&$-0.22\times10^{-5}$&11656\\ \\

\multicolumn{1}{c}{\multirow{3}{*}{$\Sigma_c^{*}\Sigma_c^{*}$}}&$1^{--}$&-157&$0.52\times10^{-5}$&0&0&4880 \\
\multicolumn{1}{c}{}&$2^{++}$&-70&0.013&$0.15\times10^{-4}$&$-0.84\times10^{-5}$&4967\\
\multicolumn{1}{c}{}&$3^{--}$&-39&0.013&$0.3\times10^{-4}$&$-0.1\times10^{-4}$&4998\\

\multicolumn{1}{c}{\multirow{3}{*}{$\Xi_c^{*}\Xi_c^{*}$}}&$1^{--}$&-165&$0.50\times10^{-5}$&0&0&5127 \\
\multicolumn{1}{c}{}&$2^{++}$&-74&0.013&$0.14\times10^{-4}$&$-0.76\times10^{-5}$&5219\\
\multicolumn{1}{c}{}&$3^{--}$&-41&0.013&$0.28\times10^{-4}$&$-0.1\times10^{-4}$&5252\\

\multicolumn{1}{c}{\multirow{3}{*}{$\Omega_c^{*}\Omega_c^{*}$}}&$1^{--}$&-173&$0.47\times10^{-5}$&0&0&5359\\
\multicolumn{1}{c}{}&$2^{++}$&-77&0.013&$0.13\times10^{-4}$&$-0.7\times10^{-5}$&5455\\
\multicolumn{1}{c}{}&$3^{--}$&-43&0.013&$0.26\times10^{-4}$&$-0.1\times10^{-4}$&5489\\ 

\multicolumn{1}{c}{\multirow{3}{*}{$\Sigma_b^{*}\Sigma_b^{*}$}}&$1^{--}$&-364&$0.23\times10^{-5}$&0&0&11296\\
\multicolumn{1}{c}{}&$2^{++}$&-162&0.013&$0.29\times10^{-5}$&$-0.16\times10^{-5}$&11499\\
\multicolumn{1}{c}{}&$3^{--}$&-91&0.013&$0.58\times10^{-5}$&$-0.23\times10^{-5}$&11570\\ 

\multicolumn{1}{c}{\multirow{3}{*}{$\Xi_b^{*}\Xi_b^{*}$}}&$1^{--}$&-372&$0.22\times10^{-5}$&0&0&11539 \\
\multicolumn{1}{c}{}&$2^{++}$&-165&0.013&$0.28\times10^{-5}$&$-0.15\times10^{-5}$&11745\\
\multicolumn{1}{c}{}&$3^{--}$&-93&0.013&$0.56\times10^{-5}$&$-0.22\times10^{-5}$&11818\\ \hline \hline
\end{tabular}
\end{small}
\end{center}
\label{t3}
\end{table}
First, the mass spectrum of the doubly charmed $qqqqcc$, $qqsscc$, $sssscc$ and bottom $qqqqbb$, $qqssbb$, $ssssbb$ dibaryons in $L=0,1,2$ states in Table\ref{t2} for $S=1$ was chosen. Since these are identical particles, we can determine the mass spectrum for each dibaryon in three orbital angular momentum states using $E_{nl}$ from Eq.\ref{e26}. We also see negligible spin terms in this table. Then, we extended our dibaryons to the other tables. From Tables \ref{t2}, \dots, and \ref{t5}, we can see the calculated mass spectrum increases with increasing quantum number and that the spin-spin effect increases slightly when these tables are considered in sequence. The spin-orbit value became more and more for each specific dibaryon in the last three tables because the $L$ and $S$ have increased. We can see spin effects on the mass spectrum of hexaquarks. $V_{SS}$ and $V_{LS}$ have a positive effect and shift up the amount of mass whereas the tensor term shifts down the mass.\\
Tables \ref{t3} and \ref{t4} have the same dibaryons for $S=1$ and $S=2$, respectively. They exhibit dibaryons with the one and two excited states. It is clearly shown that in excited dibaryon states, the mass term has a higher value than the ground states in Table \ref{t2} in the same quantum numbers.\\
In order to find out the principal quantum number effect on binding energy $E_{nl}$ in each table, we changed its amount $n=1,2,3$. It is demonstrated that $E_{nl}$ is decreasing by increasing the $n$ dramatically.\\
In Table \ref{t5}, we have the most mass spectrum and the lowest binding energy since the values of this table have been calculated for $S=3$ and $n=3$.
\begin{table}[H]
\caption{The values of hexaquarks subsystems for $S=2$, $n=2$, $L=0,1,2$, $\delta=0.5$ and $r=0.25$ are the total spin, the principal quantum number, screening and radial parameter, respectively. The masses are all in units of MeV. All data come from PDG\cite{47}.}
\begin{center}
\begin{small}
\begin{tabular}{c c c c c c c}
\hline
Hexaquark&$J^{PC}$&$E_{nl}$&$V_{SS}$&$V_{LS}$&$V_{T}$&M\\ \hline \hline
\multicolumn{1}{c}{\multirow{3}{*}{$\Sigma_c^{*}\Sigma_c$}}&$2^{-+}$&-69&$0.59\times10^{-4}$&0&0&4904 \\
\multicolumn{1}{c}{}&$3^{+-}$&-39&0.147&$0.32\times10^{-4}$&$-0.52\times10^{-4}$&4934\\
\multicolumn{1}{c}{}&$4^{-+}$&-25&0.147&$0.64\times10^{-4}$&$-0.74\times10^{-4}$&49478\\

\multicolumn{1}{c}{\multirow{3}{*}{$\Xi_c^{*}\Xi_c$}}&$2^{-+}$&-71&$0.58\times10^{-4}$&0&0&5047 \\
\multicolumn{1}{c}{}&$3^{+-}$&-40&0.147&$0.3\times10^{-4}$&$-0.49\times10^{-4}$&5078\\
\multicolumn{1}{c}{}&$4^{-+}$&-25&0.147&$0.60\times10^{-4}$&$-0.70\times10^{-4}$&5092\\	

\multicolumn{1}{c}{\multirow{3}{*}{$\Omega_c^{*}\Omega_c$}}&$2^{-+}$&-76&$0.54\times10^{-4}$&0&0&5386 \\
\multicolumn{1}{c}{}&$3^{+-}$&-43&0.147&$0.26\times10^{-4}$&$-0.43\times10^{-4}$&5419\\
\multicolumn{1}{c}{}&$4^{-+}$&-27&0.147&$0.53\times10^{-4}$&$-0.62\times10^{-4}$&5434\\	

\multicolumn{1}{c}{\multirow{3}{*}{$\Sigma_b^{*}\Sigma_b$}}&$2^{-+}$&-162&$0.25\times10^{-4}$&0&0&11480 \\
\multicolumn{1}{c}{}&$3^{+-}$&-91&0.147&$0.58\times10^{-5}$&$-0.95\times10^{-5}$&11550\\
\multicolumn{1}{c}{}&$4^{-+}$&-58&0.147&$0.12\times10^{-4}$&$-0.14\times10^{-4}$&11583\\	

\multicolumn{1}{c}{\multirow{3}{*}{$\Xi_b^{*}\Xi_b$}}&$2^{-+}$&-163&$0.25\times10^{-4}$&0&0&11584\\
\multicolumn{1}{c}{}&$3^{+-}$&-92&0.147&$0.57\times10^{-5}$&$-0.93\times10^{-5}$&11656\\
\multicolumn{1}{c}{}&$4^{-+}$&-59&0.147&$0.11\times10^{-4}$&$-0.13\times10^{-4}$&11689\\ \\

\multicolumn{1}{c}{\multirow{3}{*}{$\Sigma_c^{*}\Sigma_c^*$}}&$2^{-+}$&-70&$0.58\times10^{-4}$&0&0&4967 \\
\multicolumn{1}{c}{}&$3^{+-}$&-39&0.147&$0.31\times10^{-4}$&$-0.51\times10^{-4}$&4998\\
\multicolumn{1}{c}{}&$4^{-+}$&-25&0.147&$0.62\times10^{-4}$&$-0.72\times10^{-4}$&5012\\

\multicolumn{1}{c}{\multirow{3}{*}{$\Xi_c^{*}\Xi_c^*$}}&$2^{-+}$&-73&$0.56\times10^{-4}$&0&0&5219 \\
\multicolumn{1}{c}{}&$3^{+-}$&-41&0.147&$0.28\times10^{-4}$&$-0.46\times10^{-4}$&5252\\
\multicolumn{1}{c}{}&$4^{-+}$&-26&0.147&$0.56\times10^{-4}$&$-0.66\times10^{-4}$&5267\\	

\multicolumn{1}{c}{\multirow{3}{*}{$\Omega_c^{*}\Omega_c^*$}}&$2^{-+}$&-77&$0.53\times10^{-4}$&0&0&5455\\
\multicolumn{1}{c}{}&$3^{+-}$&-43&0.147&$0.26\times10^{-4}$&$-0.42\times10^{-4}$&5489\\
\multicolumn{1}{c}{}&$4^{-+}$&-28&0.147&$0.52\times10^{-4}$&$-0.6\times10^{-4}$&5505\\	

\multicolumn{1}{c}{\multirow{3}{*}{$\Sigma_b^{*}\Sigma_b^*$}}&$2^{-+}$&-162&$0.25\times10^{-4}$&0&0&11499\\
\multicolumn{1}{c}{}&$3^{+-}$&-91&0.147&$0.58\times10^{-5}$&$-0.94\times10^{-5}$&11570\\
\multicolumn{1}{c}{}&$4^{-+}$&-58&0.147&$0.12\times10^{-4}$&$-0.13\times10^{-4}$&11603\\	

\multicolumn{1}{c}{\multirow{3}{*}{$\Xi_b^{*}\Xi_b^*$}}&$2^{-+}$&-165&$0.25\times10^{-4}$&0&0&11746\\
\multicolumn{1}{c}{}&$3^{+-}$&-93&0.147&$0.56\times10^{-5}$&$-0.91\times10^{-5}$&11818\\
\multicolumn{1}{c}{}&$4^{-+}$&-59&0.147&$0.11\times10^{-4}$&$-0.13\times10^{-4}$&11852\\ \hline\hline
\end{tabular}
\end{small}
\end{center}
\label{t4}
\end{table}

\begin{table}[H]
\caption{The values of hexaquarks subsystems for $S=3$, $n=3$, $L=0,1,2$, $\delta=0.5$ and $r=0.25$ are the total spin, the principal quantum number, screening and radial parameter, respectively. The masses are all in units of MeV. All data come from PDG\cite{47}.  }
\begin{center}
\begin{small}
\begin{tabular}{c c c c c c c}
\hline
Hexaquark&$J^{PC}$&$E_{nl}$&$V_{SS}$&$V_{LS}$&$V_{T}$&M\\ \hline \hline
\multicolumn{1}{c}{\multirow{3}{*}{$\Sigma_c^{*}\Sigma_c^*$}}&$3^{--}$&-39&$0.18\times10^{-3}$&0&0&4998 \\
\multicolumn{1}{c}{}&$4^{++}$&-25&0.44&$0.47\times10^{-4}$&$-0.13\times10^{-3}$&5013\\
\multicolumn{1}{c}{}&$5^{--}$&-17&0.44&$0.93\times10^{-4}$&$-0.18\times10^{-3}$&5020\\

\multicolumn{1}{c}{\multirow{3}{*}{$\Xi_c^{*}\Xi_c^*$}}&$3^{--}$&-41&$0.17\times10^{-3}$&0&0&5252\\
\multicolumn{1}{c}{}&$4^{++}$&-26&0.44&$0.42\times10^{-4}$&$-0.11\times10^{-3}$&5267\\
\multicolumn{1}{c}{}&$5^{--}$&-18&0.44&$0.85\times10^{-4}$&$-0.16\times10^{-3}$&5275\\	

\multicolumn{1}{c}{\multirow{3}{*}{$\Omega_c^{*}\Omega_c^*$}}&$3^{--}$&-43&$0.16\times10^{-3}$&0&0&5489\\
\multicolumn{1}{c}{}&$4^{++}$&-28&0.44&$0.39\times10^{-4}$&$-0.1\times10^{-3}$&5505\\
\multicolumn{1}{c}{}&$5^{--}$&-19&0.44&$0.77\times10^{-4}$&$-0.15\times10^{-3}$&5513\\ \\

\multicolumn{1}{c}{\multirow{3}{*}{$\Sigma_b^{*}\Sigma_b^*$}}&$3^{--}$&-91&$0.76\times10^{-4}$&0&0&11570\\
\multicolumn{1}{c}{}&$4^{++}$&-58&0.44&$0.87\times10^{-5}$&$-0.24\times10^{-4}$&11603\\
\multicolumn{1}{c}{}&$5^{--}$&-40&0.44&$0.17\times10^{-4}$&$-0.34\times10^{-4}$&11621\\	

\multicolumn{1}{c}{\multirow{3}{*}{$\Xi_b^{*}\Xi_b^*$}}&$3^{--}$&-93&$0.74\times10^{-4}$&0&0&11818 \\
\multicolumn{1}{c}{}&$4^{++}$&-59&0.44&$0.83\times10^{-5}$&$-0.23\times10^{-4}$&11852\\
\multicolumn{1}{c}{}&$5^{--}$&-41&0.44&$0.17\times10^{-4}$&$-0.32\times10^{-4}$&11870\\ \hline\hline
\end{tabular}
\end{small}
\end{center}
\label{t5}
\end{table}
\begin{figure}
\centerline{\includegraphics[scale=0.6]{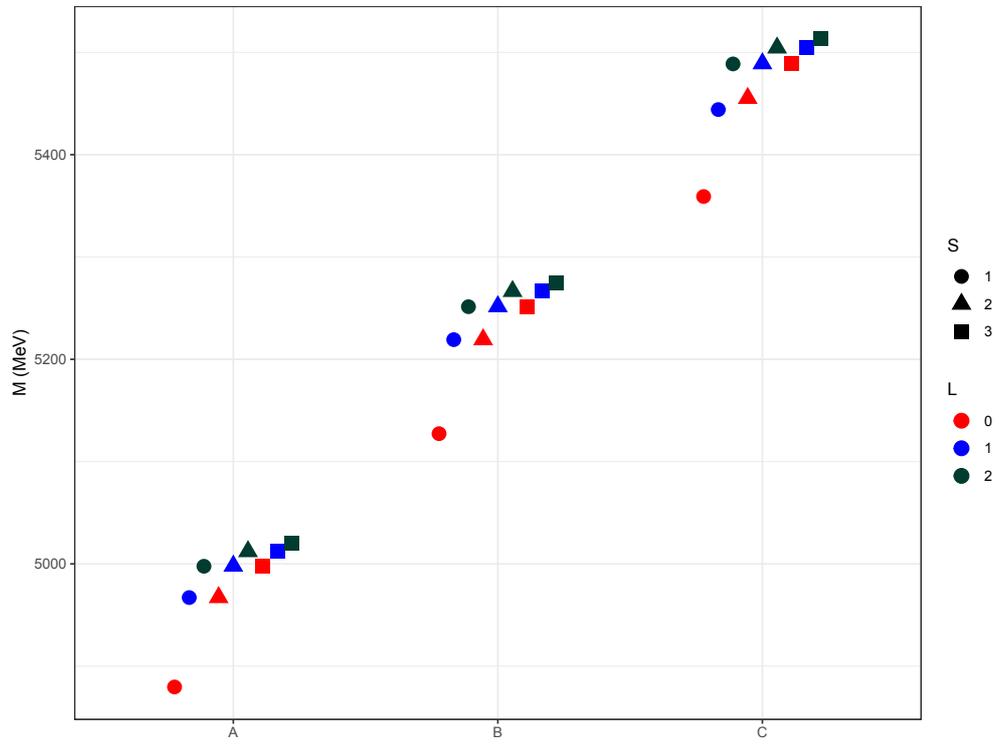}}
\caption{Calculated mass spectrum of charm dibaryons. A: $\Sigma_c^* \Sigma_c^*$, B: $\Xi_c^{*}\Xi_c^*$ and C: $\Omega_c^{*}\Omega_c^*$}
\label{fig1}
\end{figure}
\begin{figure}
\centerline{\includegraphics[scale=0.6]{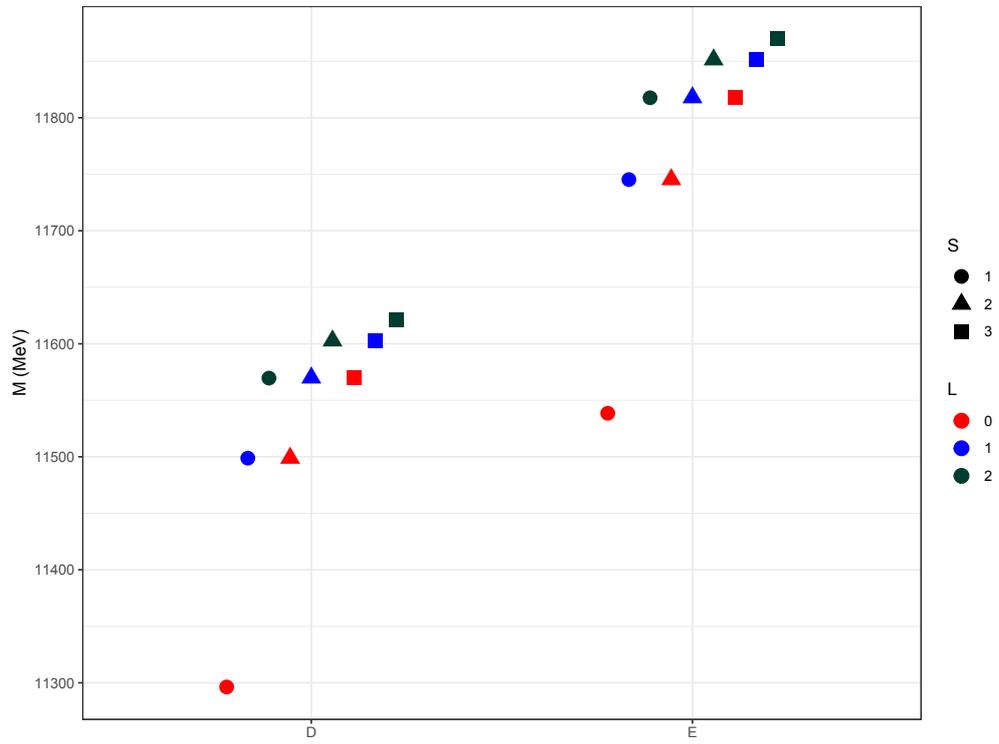}}
\caption{Calculated mass spectrum of bottom dibaryons. D: $\Sigma_b^{*}\Sigma_b^*$ and E: $\Xi_b^{*}\Xi_b^*$}
\label{fig2}
\end{figure}
\begin{figure}
\centerline{\includegraphics[scale=0.6]{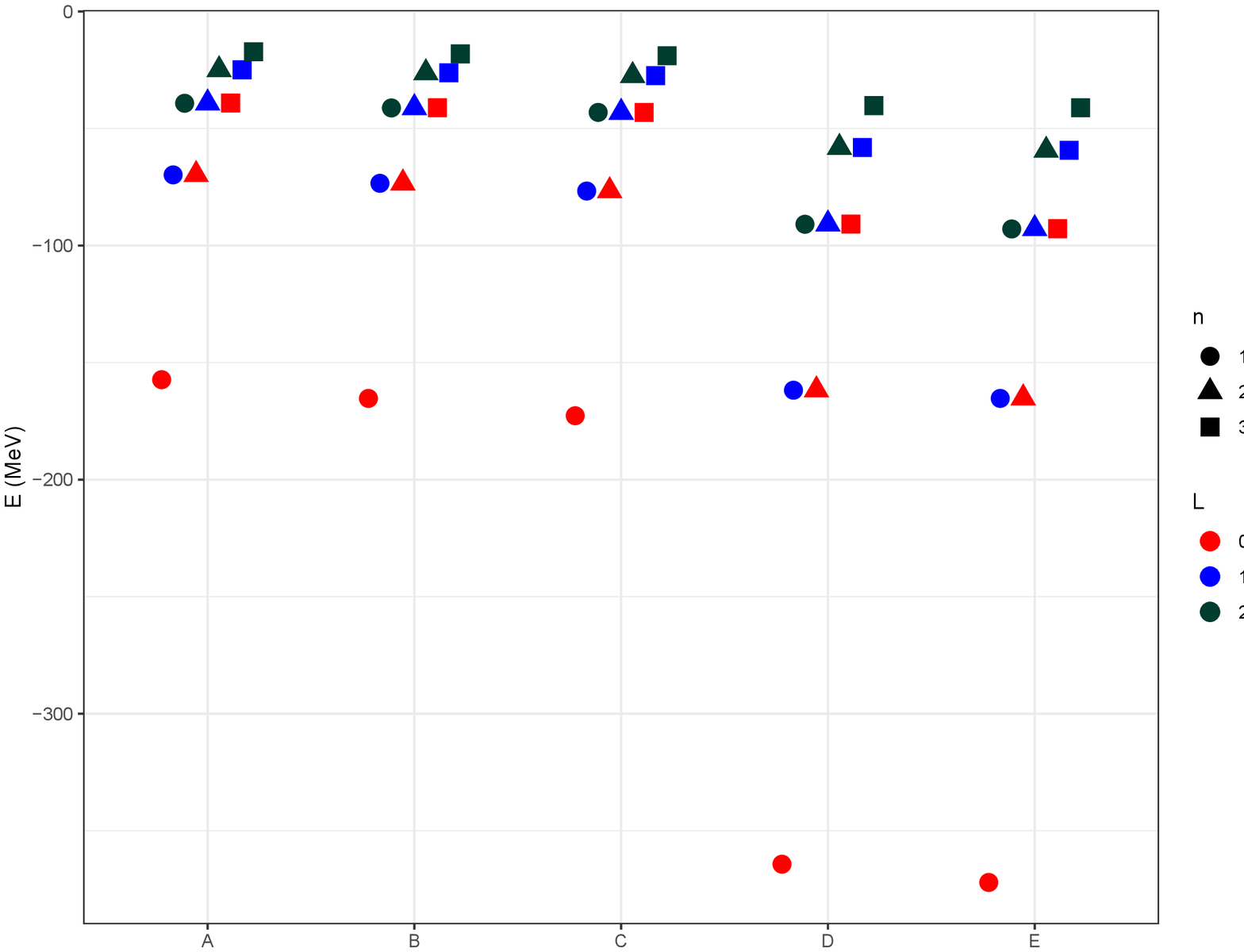}}
\caption{Calculated binding energy of charm and bottom dibaryons. A: $\Sigma_c^* \Sigma_c^*$, B: $\Xi_c^{*}\Xi_c^*$, C: $\Omega_c^{*}\Omega_c^*$ D: $\Sigma_b^{*}\Sigma_b^*$ and E: $\Xi_b^{*}\Xi_b^*$}
\label{fig3}
\end{figure}
According to the obtained mass in Table \ref{t3}, \ref{t4}, and \ref{t5} we present the mass spectrum in Figs. \ref{fig1} and \ref{fig2} for five dibaryon states. To better see the effect of the different quantum numbers on the mass of the dibaryons, we have divided the hexaquarks into two categories. Fig. \ref{fig1} is the group of hexaquarks containing two charm quarks, and Fig. \ref{fig2} is shown for hexaquarks containing two bottom quarks. For all dibaryons, we have three sets of masses in $S=1,2,3$ with $L=0,1,2$ for each spin term. The properties of these dibaryon states can be changed accordingly if the masses of these states are determined. By labeling them with red, blue, and green colors for $L=0,1,2$ and with circle, triangle, and square shapes, respectively, for $S=1,2,3$. In Figs. \ref{fig1} and \ref{fig2}, the letters of A, B, C, D, and E show $\Sigma_c^* \Sigma_c^*$, $\Xi_c^{*}\Xi_c^*$, $\Omega_c^{*}\Omega_c^*$, $\Sigma_b^{*}\Sigma_b^*$, and $\Xi_b^{*}\Xi_b^*$ hexaquarks. We can easily see that $\Sigma_c^* \Sigma_c^*$ in Fig. \ref{t1} in general has a lower mass for $L=0$ and $S=1$, and thus our results suggest that the states with lower orbital angular momentum numbers have lower masses.\\
From Fig. \ref{t3} and according to the data in Tables \ref{t5}, \ref{t4}, and \ref{t3}, we plot the binding energy $E_{nl}$ for three principal numbers $n=1,2,3$ in $L=0,1,2$. Hexaquarks in $L=2$ state have lower binding energy than the same hexaquarks in $L=1,2$ because the particles in $L=2$ are in resonance states and less bound.   
\begin{table}[H]
\caption{The values of hexaquarks subsystems for different states, compared  with other Refs. The masses are all in units of MeV.}
\begin{center}
\begin{tabular}{c c c c c c c}
\hline
Hexaquark&$J^{PC}$&Our Mass&\cite{48}&\cite{49}&\cite{50}&\cite{51}\\ \hline \hline
$\Lambda_c\Lambda_c$&$1^{--}$&4430&&&4780 \\
$\Sigma_c\Sigma_c$&$1^{--}$&4754&4930/4900&4420&&4906\\
\multicolumn{1}{c}{\multirow{2}{*}{$\Sigma_c\Sigma_c^*$}}&$1^{--}$&4817&4937/4929&4364&\\
\multicolumn{1}{c}{}&$2^{++}$&4903&4956/4942&4911&\\
\multicolumn{1}{c}{\multirow{3}{*}{$\Sigma_c^*\Sigma_c^*$}}&$1^{--}$&4880&4971/4973&4420&&5022\\
\multicolumn{1}{c}{}&$2^{++}$&4967&4946/4945&5086&&5023\\
\multicolumn{1}{c}{}&$3^{--}$&4998&4980/4991&&\\
$\Sigma_b\Sigma_b$&$1^{--}$&11258&11621&&\\
\multicolumn{1}{c}{\multirow{2}{*}{$\Sigma_b\Sigma_b^*$}}&$1^{--}$&11277&11544&&\\
\multicolumn{1}{c}{}&$2^{++}$&11479&11637&11518&\\
\multicolumn{1}{c}{\multirow{3}{*}{$\Sigma_b^*\Sigma_b^*$}}&$1^{--}$&11296&11631&&\\
\multicolumn{1}{c}{}&$2^{++}$&11499&11646&11518&\\
\multicolumn{1}{c}{}&$3^{--}$&11570&11643&&\\
\hline \hline
\end{tabular}
\end{center}
\label{t6}
\end{table}
Because of the unique and specific rescue, we chose to obtain the mass spectrum and binding energy of hexaquark states for three spin states $S=1, 2, 3$ by using all spin dependent terms for three orbital angular momenta $L=0, 1, 2$. There were not so many references to compare our results with, so in this work, we have shown a comprehensive representation that could serve as a reference for other works like ours. To compare the quantitative results of the calculation with other works with similar quark content and spin states, the results are shown in Table \ref{t6}. As we can see, the results are slightly different since all of them have their own method to obtain the mass spectrum of dibaryon states as hexaquarks.\\
In Ref. \cite{48}, they performed a systemical investigation of the low-lying doubly heavy dibaryon systems with strange $S=0$, isospin $I=0, 1, 2$, and the angular momentum $J=0, 1, 2, 3$ in the quark delocalization color screening model. They found the effect of channel-coupling cannot be neglected in the study of the multi-quark systems. In Ref. \cite{49}, the relativistic six-quark equations were constructed in the framework of the dispersion relation technique. The approximate solutions of these equations were obtained by using the method based on the extraction of leading singularities of the heavy hexaquark amplitudes,  and the poles of these amplitudes determined the masses of charm and bottom dibaryons with the isospins $I=0, 1, 2$ and the spin-parities $J^P = 0^+, 1^+, 2^+$. In Ref. \cite{50} they investigated the spectra of the prospective hidden-bottom and -charm hexaquark states with quantum numbers $J^{PC} = 0^{++}, 0^{-+}, 1^{++}, 1^{--}$ in the
framework of QCD sum rules. In Ref. \cite{51}, they considered heavy quark spin symmetry breaking and predicted several bound states of isospin $I = 0, 1, 2$ in the one-boson-exchange model. Moreover, they adopted the effective Lagrangian approach to estimate the decay widths of $\Sigma_c^*\Sigma_c^*\rightarrow \Lambda_c\Lambda_c$ and their relevant ratios via the triangle diagram mechanism.
\section{Summary}\label{sum}
Let us summarize our knowledge gained so far. So far, more and more charm-tetraquark states and pentaquark states have been discovered and confirmed by various experiments, which encourages us to investigate six-quark particles. The studies we have done in this work with the spin effects on hexaquark masses give us considerable confidence in the existence of hexaquark states. First, we introduced the AIM and extracted the energy eigenvalue equation of the corresponding radial Schr\"odinger equation based on the Hulth\'en potential for a two-body system. Meanwhile, we calculated all dependent terms, spin-spin, spin-orbit, and tensor for exhibiting the spin effects. Then we used the obtained energy eigenvalue equation for three total orbital angular momenta $L=0,1,2$ for each principal quantum number $n=1,2,3$ and calculated the energy eigenvalues for hexaquarks. We managed to gain the mass of the corresponding hexaquarks using Eq.\ref{e52}. We have listed some possible stable hexaquark states in tables\ref{t2}, \ref{t3}, \ref{t4} and \ref{t5}. To check the uncertainty of our framework, we have also consulted other works and compared them with our results. In summary, we gave a preliminary study of the mass spectra of charm and bottom hexaquark states. We hope that our study can inspire theorists and experimentalists to pay attention to these hexaquark states.
\section{Acknowledgment}
This work is supported by the university of Kashan for distinguished young scientists.

\end{document}